\documentclass[%
reprint,
superscriptaddress,
showpacs,preprintnumbers,
nofootinbib,
aps,
prb,
]{revtex4-2}

\usepackage{epsfig}
\usepackage{amsfonts}
\usepackage{amssymb}
\usepackage{bbm,bm}
\usepackage{comment}
\usepackage{amsmath}
\usepackage{cancel}
\usepackage{verbatim}
\usepackage[normalem]{ulem}
\usepackage[breaklinks, colorlinks=true, allcolors={blue}]{hyperref}
\usepackage[utf8]{inputenc}
\usepackage{graphicx}
\graphicspath{ {./figures/} }

\usepackage{lmodern}

\usepackage [english = american]{csquotes}
\MakeOuterQuote{"}
 
\newcommand\MSbar{$\overline{\text{MS}}$ } % MS-bar
\newcommand{\rmi}[1]{{\mbox{\scriptsize #1}}}
\newcommand{\rmii}[1]{{\mbox{\tiny\rm{#1}}}}

\newcommand{\g}{g}

\newcommand{\Tint}[1]{{\hbox{$\sum$}\!\!\!\!\!\!\!\int\,}_{\!\!\!\!\raise-0.9ex\hbox{$\scriptstyle{#1}$}}}
\newcommand{\Tinti}[1]{{{\Sigma}\!\!\!\!\raise0.3ex\hbox{$\int$}_\rmii{${#1}$}}}
\newcommand{\Tintip}[1]{{{\Sigma'}\!\!\!\!\!\raise0.3ex\hbox{$\int$}_\rmii{${#1}$}}}

\newcommand{\Vcl}{V_{\rmi{cl}}}

\newcommand{\te}{\textemdash}

\newcommand{\diff}{{\rm d}}
\newcommand{\ordo}[1]{\mathcal{O}({#1})}

\newcommand{\sumint}[1]{{\hbox{$\sum$}\!\!\!\!\!\!\!\int\,}_{\!\!\!\!\raise-0.9ex\hbox{$\scriptstyle{#1}$}}}
\newcommand{\suminti}[1]{{{\Sigma}\!\!\!\!\raise0.3ex\hbox{$\int$}_\rmii{${#1}$}}}

\begin{document}
\newcommand{\UPP}{\affiliation{
Department of Physics and Astronomy, Uppsala University,
Box 516, SE-751 20 Uppsala,
Sweden}}

\title{Stop comparing resummation methods}

\author{Johan L{\"o}fgren}
\email{johan.lofgren@physics.uu.se}
\UPP

\begin{abstract}
\noindent{}I argue that the consistency of any resummation method can be established if the method follows a power counting derived from a hierarchy of scales. I.e., whether it encodes a top-down effective field theory. This resolves much confusion over which resummation method to use once an approximation scheme is settled on. And if no hierarchy of scales exists, you should be wary about resumming. I give evidence from the study of phase transitions in thermal field theory, where adopting a consistent power-counting scheme and performing a strict perturbative expansion dissolves many common problems of such studies: gauge dependence, strong renormalization scale dependence, the Goldstone boson catastrophe, IR divergences, imaginary potentials, mirages (illusory barriers), perturbative breakdown, and linear terms. 
\end{abstract}

\maketitle
\section*{When perturbation theory breaks}%

\noindent{}Perturbation theory is an indispensable tool for physicists looking to connect Quantum Field Theories (QFTs) to observable quantities. A weakly coupled QFT allows for an expansion in powers of a small coupling constant, which coincides with a Feynman diagram expansion ordered by the number of loops:
\begin{equation*}
    A = A_{0}+\hbar A_{1} + \hbar^2 A_{2} + \mathellipsis
\end{equation*}
where the power of $\hbar$ denotes the loop order. This textbook treatment of perturbative expansions in QFT hence comes with its own guiding principle: order the expansion by the number of loops~\cite{Srednicki:2007qs}.

But what if the coupling is not weak? Or what if the loop integration produces a large ratio of masses? Then the loop expansion will perform poorly, and we must either (1) give up the idea of a direct perturbative treatment, (2) use different degrees of freedom, or (3) somehow improve the perturbative expansion.

Option (1) leads us to consider non-perturbative approaches, such as lattice field theory. Such approaches can be very fruitful, though they are resource-intensive and not appropriate for certain applications (such as parameter scans). A realization of option (2) can be found in the treatment of the strongly coupled low-energy limit of quantum chromodynamics, where hadronic degrees of freedom can be used instead of quarks (as in chiral perturbation theory). The success of this approach requires a sophisticated understanding of the physical theory in order to find the most useful degrees of freedom.

Option (3) can be realized by a \emph{resummation}: reorder the perturbative expansion and "re-sum" the infinite series, such that the new expansion is well-behaved. Resummation involves identifying classes of diagrams and summing them, producing a new expansion 
\begin{equation*}
    A = A'_{0}+ x A'_{1} + x^2 A'_{2} + \mathellipsis
\end{equation*}
with $x$ some new, hopefully small, parameter. Finding the correct class to resum, and how to do it, is not trivial. Also, we might be apprehensive about mixing orders in the loop expansion. After all, the loop expansion satisfies many consistency conditions that we might inadvertently break: renormalization scale independence, gauge invariance, Goldstone's theorem, \ldots And there might be several ways to reorder the expansion and resum the series. Which resummation method is correct, or failing that---which method is best?

In this paper I will focus on resummation methods, and I will argue that near all consistent resummations are implemented by integrating out heavy modes with respect to the physics you are interested in---in other words, the resummation will encode a top-down effective field theory (EFT). Because establishing an approximation scheme by choosing useful degrees of freedom and demonstrating a hierarchy of scales will enable a derivation of the correct resummation method to use, we should stop comparing resummation methods. And if there is no hierarchy of scales, we should be wary of resumming. See~\cite{Cohen:2019wxr} for a pedagogical review of EFTs.

Though this thesis applies quite generally, I will give evidence from a particular case: the study of first-order phase transitions in finite temperature field theory. A first-order phase transition is possible if the free energy of the system has two minima separated by a barrier. A continuous transition occurs if instead the minimum of the theory develops smoothly.\footnote{I will catch such continuous transitions under the umbrella term of second-order phase transitions. The difference between second- or higher-order,  or cross-over transitions, will not play a role in my argument.} 

Resummation methods used in perturbative studies of phase transitions suffer from a long list of ambiguities, such as gauge dependence, strong renormalization scale dependence, IR divergences, imaginary potentials, and more. Hence, this case serves as a difficult proving ground for a generic principle of resummation methods. 

I revisit the arguments of Arnold \& Espinosa's classic paper~\cite{Arnold:1992rz}, providing a link to modern top-down EFTs~\cite{Braaten:1995cm,Kajantie:1995dw, Cohen:2020fcu, Hirvonen:2022jba}. I emphasize that adhering to a strict power counting bypasses the long list of known issues. For simplicity, I focus on equilibrium quantities such as the critical temperature, but proper perturbative expansions are equally important when calculating dynamical quantities such as the bubble nucleation rate~\cite{Gould:2021ccf, Hirvonen:2021zej, Lofgren:2021ogg, Ekstedt:2022ceo}.

To illustrate the strength of these power-counting methods, I additionally use them to derive a novel resummation method for a class of models with radiative symmetry breaking. I conclude with a discussion of resummation methods in general, and illustrate various strengths and challenges of EFT methods.

\section*{Stop comparing resummation methods}\label{BqiN7bjeqV}
\noindent{}To set the stage for discussions of resummation methods, I want to repeat an example from~\cite{Arnold:1992rz}: that of a $\mathbb{Z}_2$-symmetric scalar field theory with classical potential
\begin{equation}
\label{eq:pure-scalar}
V_{\mathrm{cl}}(\phi)= -\frac{1}{2}\nu^2 \phi^2+\frac{1}{4!}\lambda \phi^4,
\end{equation}
where $\nu^2\ge 0$ and $\lambda \ll 1$. Though this potential exhibits spontaneous symmetry breaking, it is \textit{a priori} not clear if a first-order phase transition occurs at some high temperature.

Using the formalism of finite temperature field theory (see~\cite{Kapusta:2006pm, Laine:2016hma} for pedagogical introductions), we can derive the perturbative corrections to this potential when the scalar field is coupled to a thermal bath of temperature $T$. In particular, if the scalar's field-dependent mass is  $m^2(\phi) \ll T^2$, the 1-loop effective potential is
\begin{equation}
\label{W57Y2gVVtP}
V_{1}(\phi)=\frac{1}{24} m^2(\phi) T^2 -\frac{1}{12 \pi}\left(m^2(\phi)\right)^{3/2}T+\mathellipsis
\end{equation}
Using the square mass $m^2(\phi)=V_{\mathrm{cl}}''(\phi)$ gives the leading contribution
\begin{equation}
V_{1}(\phi)=\frac{1}{48}\lambda T^2 \phi^2,
\end{equation}
where I discarded field-independent terms. Now, for a phase-transition to occur, this term must be of similar size as the tree-level potential. This indicates that we should add its contribution to Equation~\eqref{eq:pure-scalar},
\begin{equation}
\label{eq:ScalarLO}
V_{\mathrm{LO}}(\phi)=\frac{1}{2}\left( -\nu^2+\frac{1}{24}\lambda T^2\right)\phi^2+\frac{1}{4!}\lambda \phi^4,
\end{equation}
and instead consider $V_{\mathrm{LO}}(\phi)$ as the leading-order potential.

This means that the high-temperature $\phi^2$-coefficient is
\begin{equation}\label{eq:nueff}
\nu_{\mathrm{eff}}^2=-\nu^2 + \frac{1}{24}\lambda T^2.
\end{equation} 
If we consider a cooling plasma in which $T$ is decreasing, a second-order phase transition would occur if this coefficient changes sign before another minimum is generated. Hence, we can find the second-order transition temperature $T_0$ by solving $\left.\nu_{\mathrm{eff}}^2\right|_{T_0}=0$, which gives
\begin{equation}
    T_0 = \sqrt{\frac{24\nu^2}{\lambda}}.
\end{equation}

We can also find the effective---resummed---square mass from the leading-order potential,
\begin{equation}
    m^2_{\mathrm{eff}}(\phi)=V_{\mathrm{LO}}''(\phi)=-\nu_{\mathrm{eff}}^2+\frac{1}{2}\lambda \phi^2.
\end{equation}
This resummation method is now known as Arnold-Espinosa resummation. It is implemented by resumming the masses of the light modes by integrating out the heavy modes with momenta of order $T$. The loops come with quadratic UV divergences and yield $T^2$; with the coupling $\lambda$ the contributions are of order $\lambda T^2$~\cite[p. 6-7]{Arnold:1992rz}.\footnote{The page numbers apply to the ArXiv version of~\cite{Arnold:1992rz}.} The effect of the resummation when calculating the 1-loop potential is realized by adding a \emph{ring-improved} term,
\begin{equation}\label{eq:ringpotential}
    V_{\mathrm{ring}}(\phi)=-\frac{T}{12 \pi} \left[\left(m_{\mathrm{eff}}^2(\phi)\right)^{3/2}-\left(m^2(\phi)\right)^{3/2}\right],
\end{equation}
effectively replacing the mass of the particle in the contribution from light modes to the effective potential.

Adding this contribution to the potential in equation~\eqref{eq:ScalarLO} gives
\begin{equation}
V(\phi)=\frac{1}{2}\nu_{\mathrm{eff}}^2\phi^2 -\frac{T}{12\pi}\left(-\nu_{\mathrm{eff}}^2+\frac{1}{2}\lambda \phi^2\right)^{3/2}+\frac{1}{4!}\lambda \phi^4,
\end{equation}
and we might be tempted to interpret this as the new leading-order potential. I will return to whether this is sensible, and whether this theory has a first- or second-order phase transition, in a later section.

The resummation method suggested by Arnold \& Espinosa can more formally be realized as constructing a three-dimensional Euclidean EFT in terms of Matsubara zero modes~\cite{Ginsparg:1980ef,  Appelquist:1981vg}. This EFT will have Arnold-Espinosa resummation as a leading-order prediction. The process of constructing the 3D EFT is known as \emph{dimensional reduction}, and it enables systematic extension of the resummation method to higher orders~\cite{Braaten:1995cm,Kajantie:1995dw}. 

When the high-temperature expansion applies, Arnold-Espinosa resummation (dimensional reduction) is the best approach. But what if the high-temperature expansion does not apply? Many then turn to Parwani resummation, where thermal masses are inserted in the full effective potential~\cite{Parwani:1991gq}. This can be implemented with the help of a "thermal counterterm," effectively subtracting off terms that would otherwise be double-counted.

I will write the contribution to the \MSbar{}-renormalized 1-loop potential for a bosonic degree of freedom with square mass $m^2$ at temperature $T$ as~\cite{Dolan:1973qd}
\begin{align}
    J(m^2)&=J_0(m^2)+\frac{T^4}{2 \pi^2} J_B(\frac{m^2}{T^2}),\\
    J_0(m^2)&=\frac{m^4}{64 \pi^2} \left(\log \left[\frac{m^2}{\mu^2}\right]-\frac{3}{2}\right),\label{eq:1loopZerotemp}\\
    J_B(y^2)&=\int \diff x ~ x^2 \log\left[1-\exp\left\{-\sqrt{x^2+y^2}\right\} \right],
\end{align}
with $\mu$ the \MSbar{} scale.\footnote{The constant $-3/2$ in $J_0$ is slightly different for a vector boson~\cite{Coleman:1973jx}. Here I also ignore the analogous fermionic functions.} With this notation, we can implement Parwani resummation of the 1-loop potential by adding the term
\begin{equation}
    V_{\mathrm{Parwani}}=J(m^2_{\mathrm{eff}})-\sum_{i=0}^{\infty}(\delta m_T^2)^i J^{(i)}(m^2),
\end{equation}
where $\delta m_T^2$ is the thermal counterterm that implements the resummation, and $J^{(i)}(m^2)$ is the $i$:th derivative of $J(m^2)$ with respect to $m^2$. Each power of $(\delta m_T^2)$ raises the loop-order: the subtracted terms should be sorted into their appropriate loop orders to cancel terms and prevent double-counting. Of course, if any other loop functions are resummed then the corresponding terms should be subtracted similarly.

Parwani resummation does not depend directly on the high-temperature expansion, and as long as one is consistent in subtracting the diagrams at each loop order, the resummation method should not introduce any problems. Hence, we might draw the conclusion that the Parwani method is a safer bet when the high-temperature expansion does not apply---as is often the case in phenomenological models with many particles of varying masses. (See the subsection on linear terms for an example of what can go wrong if one is not consistent.)

On the other hand, Arnold-Espinosa resummation has conceptual clarity: only the modes which require resummation are resummed. Double-counting is never even an issue.

Because the two methods have their respective strengths, they are compared against each other in numerical studies (see e.g.~\cite{Kainulainen:2019kyp, Athron:2022jyi}, and the studies cited within), and discrepancies between the two methods are often found. Yet such comparisons miss the point.

In fact, Arnold \& Espinosa originally compared their resummation method to that of Parwani, concluding that the methods give equivalent results~\cite[p. 25]{Arnold:1992rz}. But if this is the case, how come the previously mentioned studies keep finding that the methods give different results? The reason is that the conclusion of Arnold \& Espinosa rests on a few assumptions. First, that the high-temperature expansion applies. Second, the existence of a consistent power counting---an expansion in a small parameter.

Any perturbative study of a phase transition in which the high-temperature expansion does not apply, or if it is not an expansion in a small parameter, will show a difference between these two methods. This could give the illusion that the two methods give different results, but a difference between the two methods simply reveals that the perturbative expansion, one way or another, is not working.

This is why I stress that the true lesson of Arnold \& Espinosa's paper is not their resummation method---which now is supplanted by dimensional reduction anyway---but the principle of using strict perturbative expansions. 

Arnold \& Espinosa also emphasize that consistency of any one resummation method requires the protection of a hierarchy of scales, claiming that a mass-resummation is only reasonable when the self-energy is not small compared to the inverse propagator. And only then can the momentum dependent self-energy $\Pi(p^2)$ be approximated by $\Pi(0)$~\cite[p. 20]{Arnold:1992rz}.

To see this, consider a propagator $D(p^2)$ improved by including the momentum-dependent self-energy $\Pi(p^2)$, with $\hbar \ll 1$ as a loop-counting parameter,
\begin{align}
    D(p^2)&=\frac{1}{p^2+m^2+\Pi(p^2)},\\
    \Pi(p^2)&=\hbar\Pi_1(p^2)+\hbar^2\Pi_2(p^2)+\mathellipsis.
\end{align}
By expanding the self-energy in powers of momentum, 
\begin{equation}
    \Pi(p^2)=\Pi(0)+p^2\Pi'(0)+\mathellipsis,
\end{equation}
we can see that we need
\begin{equation}
    p^2\Pi'(0) \ll \Pi(0)
\end{equation}
for the momentum-expansion to apply.
Now we can consider a generic example, in which the self-energy contains contributions from a heavier particle with mass $M^2 \sim m^2/\hbar $. Then
\begin{equation}
    x\Pi_1(p^2)\sim \hbar M^2 + \hbar p^2.
\end{equation}
In this case, with $p^2 \sim m^2$, we have
\begin{align}
    m^2 + \Pi(p^2)&= m^2 + \hbar \left( \Pi_1(0) + p^2\Pi_1'(0) \right) + \ordo{\hbar^2}\nonumber\\
    &\sim m^2 + \hbar M^2 + \hbar p^2 + \ordo{\hbar^2}.
\end{align}
So the hierarchy between $m^2$ and $M^2$ tells us that we can define a new effective mass
\begin{equation}
    m^2_{\mathrm{eff}}\sim m^2 + \hbar M^2,
\end{equation}
and that the momentum dependence of the self-energy can be neglected for this purpose. At higher orders, the momentum dependence can become important and can be included through higher-order derivative operators in the action.

If there is no hierarchy of scales, then we are not justified in simply resumming the mass: the whole self-energy is needed.\footnote{As Arnold \& Espinosa note in their appendix C~\cite[p. 58]{Arnold:1992rz}, this poses a challenge for super-daisy resummations~\cite{Brahm:1991nh} and partial dressing resummation~\cite{Boyd:1993tz}, in which a gap equation is solved to find the resummed mass. This remains a challenge for modern variations of this approach, as in~\cite{Curtin:2016urg, Curtin:2022ovx}.}

\section*{Establish a hierarchy of scales}
\noindent{}In this section, I discuss four different scale hierarchies and show how EFTs constructed from these hierarchies can implement resummations.
\subsection*{Hierarchy I}
\noindent{}To begin with, consider a theory with one mass scale $m$ at temperature $T$ such that the hierarchy
\begin{equation}\label{}
    m \ll \pi T
\end{equation}
holds. For concreteness, we can consider the pure scalar theory defined in equation~\eqref{eq:pure-scalar}, with $-\nu^2 \sim \lambda T^2$. The hierarchy implies that we can integrate out the heavy modes, to get a resummed theory of light modes with $\phi^2$-coefficient $\nu_{\mathrm{eff}}^2$ as given in equation~\eqref{eq:nueff}. A cheap way to implement this resummation is to use the ring-improved potential of equation~\eqref{eq:ringpotential} defined by Arnold \& Espinosa~\cite{Arnold:1992rz}.

The more systematic version of this resummation method is called \emph{dimensional reduction}~\cite{Ginsparg:1980ef,  Appelquist:1981vg, Braaten:1995cm, Kajantie:1995dw}. To give a brief motivation: the equilibrium quantities of a finite temperature field theory can be studied in an imaginary-time formalism. Bosonic fields are periodic over this "time" direction, with period $\sim 1/T$ where $T$ is the temperature. This allows the fields to be decomposed into Matsubara modes $\varphi_n$ with masses $m^2+(2 \pi n T)^2$. The modes with $n\neq 0$ are analogous to heavy particles, and can be integrated out with standard field-theory techniques. See~\cite{Gould:2021dzl, Niemi:2021qvp, Schicho:2021gca} for modern and pedagogical reviews of this concept; see~\cite{Ekstedt:2022bff} for software which automatizes the matching procedure for generic models.

We are then left with a three-dimensional Euclidean theory of zero-modes. The potential is now
\begin{equation}\label{eq:3dscalar}
    V(\phi_3)=\frac{1}{2}m_3^2 \phi_3 ^2+\frac{1}{4!}\lambda_3 \phi_3^4, 
\end{equation}
with 3D quantities (note their mass dimensions)
\begin{align}
    \phi_3^2 &= \frac{\phi^2}{T}+\mathellipsis,\\
    m_3^2 &=-\nu^2+\frac{1}{24}\lambda T^2 + \mathellipsis,\\
    \lambda_3 &= \lambda T +\mathellipsis.
\end{align}
Note that the $\phi_3^2$ coefficient $m_3^2$ corresponds to the resummed $\nu_{\mathrm{eff}}^2$ in equation~\eqref{eq:nueff}. Here the ellipses hide higher-order corrections which can be found by performing the matching to higher orders.

Now we can face whether the pure scalar theory has a first- or second-order phase transition. For a first-order phase transition to occur, a barrier needs to develop in order to have two separate minima. The potential in equation~\eqref{eq:3dscalar} does not have a barrier between the symmetric minimum and the broken minimum for any values of the coefficients. Can a barrier arise from 1-loop corrections?

The contribution to the 1-loop potential in the 3D EFT for a field of square mass $M^2(\phi_3)$ is
\begin{equation}
    f_3(M^2)=-\frac{1}{12\pi}\left(M^2\right)^{3/2}.
\end{equation}
Adding this to the tree-level potential gives
\begin{equation}\label{eq:scalar-barrier-3d}
    V_{\mathrm{LO}}(\phi_3)=\frac{1}{2}m_3^2 \phi_3 ^2-\frac{1}{12\pi}\left(m_3^2+\frac{1}{2}\lambda_3 \phi_3^2\right)^{3/2}+\frac{1}{4!}\lambda_3 \phi_3^4, 
\end{equation}
and we assume for now that this is the new leading-order potential. The new term can give rise to a barrier if $m_3^2$ is small such that a cubic term $\phi_3^3$ is generated.

To understand whether this potential makes sense as a leading-order expression, we should perform a power counting. Balancing the terms in equation~\eqref{eq:scalar-barrier-3d} gives
\begin{equation}\label{eq:scalar-barrier-pc-3d}
    \phi_3 \sim \sqrt{\lambda_3}, m_3^2 \sim \lambda_3^2 \implies M(\phi_3)\sim \lambda_3.
\end{equation}
Even though \emph{a priori} this counting seems innocuous---the high-temperature expansion clearly applies, and $m_3^2$ is indeed small---there is a problem here. Because this theory only has one coupling constant $\lambda_3$ (with mass-dimension $1$) and one effective mass $M(\phi_3)$ (derived from the leading-order potential), each time we go up in loop order we must add a factor of $\lambda_3$ whereby the dimensions force us to remove one factor of $M(\phi_3)$. The loop expansion of the effective potential then shows the sequence
\begin{equation}
    \lambda_3^{-1}M^4, M^3, \lambda_3 M^2, \lambda_3^2 M, \lambda_3^3, \mathellipsis
\end{equation}
for loop orders zero, one, two, three, four, \ldots. And from the power counting in equation~\eqref{eq:scalar-barrier-pc-3d} we know $M\sim \lambda_3$. Hence, all loop orders contribute at the same order in perturbation theory: $\lambda_3^3$.

More formally, rescaling the field as $\phi_3 \rightarrow \sqrt{m_3} \phi_3$, the momenta as $p_i \rightarrow m_3 p_i$  and defining $x= \lambda_3/m_3$ yields the dimensionless potential
\begin{equation}
        \frac{V(\phi_3)}{m_3^3}=\frac{1}{2} \phi_3 ^2+\frac{1}{4!}x \phi_3^4.
\end{equation}
The only coupling constant of this theory is $x$, and the power counting gives $x\sim 1$. This is not an expansion in a small parameter, and our conclusions based on it cannot be trusted. Its barriers are only mirages.

A more intuitive formulation: in order for the 1-loop correction to change the shape of the potential, it needs to be big enough to affect the classical potential. We need something heavy to amplify the 1-loop potential, but we only have a single scalar field\te and it can not be heavier than itself.

\subsection*{Hierarchy II}
\noindent{}Next I will consider a scale hierarchy with an intermediate scale,
\begin{equation}
    m \ll M \ll \pi T.
\end{equation}
This scale hierarchy offers a rich set of possibilities. One example is that of a gauge theory at high temperature. We can study the Abelian Higgs model---a complex scalar charged under a $\mathrm{U}(1)$ gauge field---by integrating out high-energy modes in two steps. First a dimensional reduction is performed, and then the gauge boson is integrated out~\cite[p. 30]{Arnold:1992rz}. Note that the procedure requires integrating out a field whose mass and couplings depend on the background field of a lighter scalar. This will yield a non-polynomial effective action~\cite{Weinberg:1992ds, Hirvonen:2022jba}; the construction of such effective actions is also known as functional matching~\cite{Cohen:2020fcu}. See also~\cite{Jakovac:1995ra, Jakovac:1996kv} for early attempts at constructing non-polynomial effective actions for studying phase transitions in gauge theories. Following the demonstration in~\cite{Hirvonen:2022jba}, formally the method entails rewriting the partition function
\begin{equation}
    Z = \int \mathcal{D}\Phi e^{-S[\Phi]}
\end{equation}
by separating the UV and IR modes of the fields: $\mathcal{D}\Phi=\mathcal{D}\Phi^{\textsc{uv}}\mathcal{D}\Phi^{\textsc{ir}}$, and performing the integral over the UV modes:
\begin{align} 
    Z &= \int \mathcal{D}\Phi^{\textsc{ir}} e^{-S_{\mathrm{eff}}[\Phi^{\textsc{ir}]}},\\
    S_{\mathrm{eff}}[\Phi^{\textsc{ir}}]&= -\log\int\mathcal{D}\Phi^{\textsc{uv}}e^{-S[\Phi^{\textsc{ir}}+\Phi^{\textsc{uv}}]}.
\end{align}
See~\cite{Hirvonen:2022jba} for a detailed account of how to perform this integral.

Denoting the gauge coupling by $g$, we get the leading-order potential and parameters
\begin{align}\label{eq:potential-ah-3d}
    V_{\mathrm{LO}}(\phi_3)&=\frac{1}{2}m_3^2 \phi_3 ^2-\frac{1}{12\pi}\left(g_3^2 \phi_3^2\right)^{3/2}+\frac{1}{4!}\lambda_3 \phi_3^4, \\
    m_3^2&=m^2+\frac{g^2}{12}T^2+\frac{1}{18}\lambda T^2 + \mathellipsis, \label{eq:AHMatching}\\
    \lambda_3&=\lambda T + \mathellipsis,\\
     g_3^2&=g^2 T + \mathellipsis.
\end{align}
We see that the leading-order potential has a barrier via the $\phi^3$ term. To ensure that this is a well-formed leading-order expression, we repeat the exercise of balancing terms. The result is~\cite[p. 9-10]{Arnold:1992rz}
\begin{equation}
    m_3^2\sim \frac{g_3^6}{\lambda_3}, \phi_3\sim \frac{g_3^3}{\lambda_3}.
\end{equation}
But we have the additional constraint (from the assumed scale hierarchy) that the gauge boson is heavier than the scalar, $\g_3 \phi_3 / m_3 \sim g_3/\sqrt{\lambda_3}\ll 1$. A simple realization of this hierarchy is to assume that $\lambda \sim g^3$ (in contrast with $\lambda \sim g^2$, which is the standard assumption in loop expansions). So the potential in equation~\eqref{eq:potential-ah-3d} is actually a well-behaved leading-order expression.

By deriving the masses of the scalars from this potential, it is possible to extend previous studies~\cite{Karjalainen:1996rk,Kajantie:1997hn} to study phase transitions in this theory accurately and consistently~\cite{Ekstedt:2022ceo,Hirvonen:2021zej,Lofgren:2021ogg,Ekstedt:2022zro}.
\subsection*{Hierarchy III}
\noindent{}We can also imagine another hierarchy in which the heavy field is so heavy it is not excited by the temperature $T$,
\begin{equation}\label{eq:hierarchy-heavy-field}
    m \ll \pi T \ll M.
\end{equation}
In this case we should first integrate out all the modes of the heavy field, and then integrate out the non-zero Matsubara modes of the light field. See~\cite{Hirvonen:2022jba} for an example.
\subsection*{Hierarchy IV}
\noindent{}Consider a heavy field at a scale close to the temperature $T$,
\begin{equation}\label{eq:hierarchy-heavy-field-at-T}
    m \ll M \sim \pi T.
\end{equation}
In this case, the high-temperature expansion does not apply to the field of mass $M$. But neither can the temperature be neglected when integrating it out, as in hierarchy III. However, the high-temperature expansion still applies for the light field of mass $m$. There should still exist a 3D EFT for the zero-mode of the light field. This method of "partial dimensional reduction" is not widely studied, but see~\cite{Laine:2000kv,Laine:2000rm,Brauner:2016fla,Laine:2019uua} for a few studies.

Here I want to highlight another example: a variant of the Coleman-Weinberg (CW) model~\cite{Coleman:1973jx} as studied in~\cite{Metaxas:1995ab}. This model features \emph{radiative symmetry breaking}: there is no symmetry breaking at tree-level, but there is at 1-loop level. This is the Abelian Higgs model with a small and positive mass term:
\begin{equation}
    \Vcl(\phi)=\frac{1}{2}m^2 \phi^2+\frac{1}{4!}\lambda\phi^4,
\end{equation}
This potential is of comparable size to the 1-loop contribution of the gauge boson when $\lambda \sim g^4$ and $m^2\sim g^4 \sigma^2$ with $\sigma$ a characteristic size of the VEV. 

There is a clear hierarchy of scales: the gauge boson is heavier than the scalars and can be integrated out. This gives a modified potential
\begin{equation}
    V_{\mathrm{LO}}(\phi)=\frac{1}{2}m^2 \phi^2+\frac{1}{4!}\lambda\phi^4+3J_{\mathrm{CW}}(g^2 \phi^2),
\end{equation}
where $J_{\mathrm{CW}}$ is the same as $J_0$ of equation~\eqref{eq:1loopZerotemp} but with $-3/2$ replaced by $-5/6$. This potential has a non-zero minimum, and it is from this potential which we should find the scalar masses,
\begin{align}
    m_H^2(\phi)&=V''_{\mathrm{LO}}(\phi),\label{eq:resummation-Higgs}\\
    m_G^2(\phi)&=\frac{V'_{\mathrm{LO}}(\phi)}{\phi}.\label{eq:resummation-Goldstone}
\end{align}
This is a consistent mass resummation derived from power-counting rules, with a hierarchy of scales that protects it from double-counting diagrams and other issues.

This model has two different minima with a barrier in between---could there be a first-order phase transition between them? To approach this question, we can assume that the high-temperature expansion applies, and that a $\phi^3$ barrier is induced:
\begin{align}
    V_{\mathrm{LO}}(\phi)&=\frac{1}{2} m^2_{\mathrm{eff}} \phi^2+\frac{1}{4!}\lambda\phi^4 -\frac{3}{12 \pi}T g^3 \phi^3\nonumber\\
    &+ 3J_{\mathrm{CW}}(g^2 \phi^2),
\end{align}
but by balancing the powers of this expression we find
\begin{equation}\label{eq:CW-temp-pc}
    g \phi \sim T, m^2_{\mathrm{eff}}\sim g^2 T^2.
\end{equation}
Which implies that the high-temperature expansion does not apply, as then the gauge boson mass goes as $g \phi \sim T$~\cite{Kierkla:2022odc}.

But really we are asking for too much: we do not need the high-temperature expansion to apply to the gauge field. This model already has two minima, we do not need the temperature effects to create a barrier. We only need it to shift the energy of the different minima such that a phase transition can occur~\cite{Linde:1978px, Ginsparg:1980ef}; the resulting potential reads
\begin{align}
    V_{\mathrm{LO}}(\phi)&=\frac{1}{2} m^2_{\mathrm{eff}} \phi^2+\frac{1}{4!}\lambda\phi^4 \nonumber\\
    &+3J_{\mathrm{CW}}(g^2 \phi^2) + 3 \frac{T^4}{2 \pi^2} J_B(g^2 \phi^2).
\end{align}
Balancing the powers here gives the same counting as in equation~\eqref{eq:CW-temp-pc}.

But the question remains what to do with the scalar field. After all, it is this field which will potentially undergo a transition.

Because the high-temperature expansion still applies to the scalar field, we should treat it using a 3D EFT as before. To reach this EFT, we integrate out the heavy modes of the scalars: the non-zero Matsubara modes and the high-momentum modes of the Matsubara zero-mode. At the same time, we also integrate out \emph{all} modes of the vector. What we end up with is a Euclidean 3D EFT with potential
\begin{align}
    V_{\mathrm{LO}}(\phi_3)&=\frac{1}{2} m^2_{3} \phi_3^2+\frac{1}{4!}\lambda_3\phi_3^4 \nonumber\\
    &+3\frac{J_{\mathrm{CW}}}{T}(g^2 \phi^2) + 3 \frac{T^3}{2 \pi^2} J_B(g^2 \phi^2).\label{eq:CW-highT-effpot}
\end{align}
The 3D parameters are determined by matching with the 4D theory. In this case we find
\begin{align}\label{eq:CWMatching}
    m_3^2&=m^2+\frac{1}{18}\lambda T^2 + \mathellipsis,\\
    \lambda_3&=\lambda T + \mathellipsis,\\
    \phi_3^2&=\frac{\phi^2}{T}+ \mathellipsis.
\end{align}
The difference between these expressions and those of regular dimensional reduction of the Abelian Higgs model is that here the non-zero Matsubara modes of the vector boson do not contribute directly to the Wilson coefficients (compare equations~\eqref{eq:CWMatching} and~\eqref{eq:AHMatching} and note the missing $g^2 T^2$ term). Instead, the vector modes contribute through the non-polynomial term in the effective potential. This contribution will in the end propagate to the mass of the scalar.

To find the correct resummation to use in this theory, we use the same derivatives as in equations~\eqref{eq:resummation-Higgs} and~\eqref{eq:resummation-Goldstone}, but with the potential given by equation~\eqref{eq:CW-highT-effpot}. I emphasize that this resummation contains parts that are not utilizing the high-temperature expansion. And yet this should be a wholly consistent resummation. Furthermore, because the high temperature expansion applies to the scalar field which undergoes the phase transition, the machinery of thermal escape (tunneling at finite temperature)~\cite{Gould:2021ccf, Ekstedt:2021kyx, Ekstedt:2022tqk} should apply and all the usual formulas carry over.\footnote{A funny corollary of this is that a CW-like $\mathrm{SU(2)}$ gauge theory would automatically suppress sphaleron transitions after the phase transition. The power counting in equation~\eqref{eq:CW-temp-pc} implies that the gauge field is not excited at the phase transition temperature, since it is too heavy. The suppression of thermal sphalerons in models with radiative symmetry breaking hence make them natural candidates for electroweak baryogenesis (see~\cite{Prokopec:2018tnq} for a numerical study in agreement with this claim).} Though, a detailed analysis of the power-counting scheme and its convergence is warranted.

This expansion would also work for the more typical CW model without a positive mass, if there are other fields in the theory with masses $\ll T$. Even though the tree-level potential then does not have a barrier, one is generated at finite temperature because the other light fields contribute to $m^2_{\mathrm{eff}}$, giving a positive $m^2_{\mathrm{eff}}\sim g^2 T^2$. This power counting may hence be of use in modifications of the standard model of particle physics with radiative symmetry breaking, such as the one studied in~\cite{Kierkla:2022odc}.
\section*{Dissolve illusory problems}
\noindent{}Studies of the electroweak phase transition have a long laundry list of problems: gauge dependence, strong renormalization scale dependence, the Goldstone boson catastrophe, IR divergences, imaginary potentials, mirages, perturbative breakdown, resummation method dependence, and linear terms. Many of these problems were recently studied in~\cite{Athron:2022jyi}, where it was shown that some of them can yield big quantitative and qualitative uncertainties.

In this section, I review these problems and argue that they are dissolved if one uses a consistent and strict perturbative expansion.
\subsection*{Gauge dependence}
\noindent{}The gauge dependence of the effective action, and in particular the effective potential, is well-known and captured in the famous Nielsen identities~\cite{Nielsen:1975fs,Fukuda:1975di}. Essentially, the effective potential is only gauge-invariant when evaluated at an extremum---at a physical point. But in perturbation theory there are implementation details: to get a gauge-independent result we must use a \emph{strict perturbative expansion}. So if the effective potential is expanded as
\begin{equation}
    V=V_0+x V_1+x^2 V_2+\mathellipsis,
\end{equation}
then we must find the extrema perturbatively,
\begin{equation}
    \phi=\phi_0+x\phi_1+x^2 \phi_2 +\mathellipsis,
\end{equation}
by inserting this expansion of $\phi$ into the expansion of $V$, and extremizing the potential order-by-order:
\begin{align}
    V'{\large{|}}_\phi=0 \implies V'_0\large{|}_{\phi_0}&=0,\\
     \phi_1&=-\left.\frac{V'_1}{V''_0}\right|_{\phi_0},\\
    &~~\!\vdots\nonumber
\end{align}
This expansion is sometimes called a tadpole expansion, since it effectively reinserts scalar tadpoles into the 1PI diagrams of the effective potential~\cite{Fukuda:1975di}.

Though it is well-established that the strict expansion above gives gauge-independent results, there has remained some confusion if it can also give accurate results. The strict expansion was popularized in~\cite{Patel:2011th} under the name $\hbar$-expansion (now sometimes called the PRM method). The authors then expressed concern that the $\hbar$-expansion required a strict loop counting, while any resummation necessarily mixes loop orders. 

The way out of this dilemma is to realize that though a strict expansion is necessary, it does not have to be a loop expansion. All that is required is that the perturbative expansion is performed using a consistent power counting~\cite{Ekstedt:2020abj}. In~\cite{Ekstedt:2022zro} it was shown that such a strict expansion works if the expansion parameter $x$ is small. There is no conflict between gauge-independence and accurate results~\cite[p. 26]{Arnold:1992rz}.\footnote{Unfortunately, this is not reflected in the wider literature. There is much confusion, as is evident by sampling the papers citing~\cite{Patel:2011th}.}

\subsection*{Strong renormalization scale dependence}
\noindent{}There are many studies that demonstrate a strong renormalization scale dependence in perturbative calculations of phase transition quantities~\cite{Croon:2020cgk,Gould:2021oba, Athron:2022jyi}. The problem is that resummations mix loop orders, which messes up the ordinary cancellation between implicit running of parameters and explicit running of loop-functions. As the thermal masses arise at one loop, their running must be cancelled by the next loop order: two loops.

As such, the solution is to use dimensional reduction and calculate up to two-loop order. Constructing the 3D EFT consistently resums large contributions, and the running within this EFT is tame.
\subsection*{The Goldstone boson catastrophe}
\noindent{}To understand a possible source of IR-divergences we can consider the form of the zero-temperature 1-loop function $J_0(m^2)$ and its second derivative with respect to $m^2$ in the small $m^2$ limit,
\begin{equation}
    J_0(m^2) \sim (m^2)^2 \log m^2 \implies J_0''(m^2)\sim \log m^2.
\end{equation}
The second derivative of this function diverges in the $m^2\rightarrow 0$ limit. This divergence indicates two related problems.

The first problem can be seen if we think of these derivatives as insertions of interactions. Then the divergence of $J_0''(m^2)$ implies that the 3-loop potential will diverge in the same limit. This is known as the Goldstone boson catastrophe, since the Goldstones are massless in the broken minimum. The literature implies a resummation is necessary to cure this divergence~\cite{Martin:2014bca, Elias-Miro:2014pca, Espinosa:2016uaw}. But this is not always true. In fact, a resummation is not necessary to remove the divergence in a regular loop expansion~\cite{Ekstedt:2018ftj}. It is enough to simply perform a strict expansion: any IR divergences of the potential are cancelled by corresponding divergences in the tadpoles. However, if a modified power counting is used---i.e. when there is a hierarchy of scales---resummation becomes necessary again. 

The second related problem arises if one attempts to use a mass-dependent renormalization scheme and tries to match the measured masses of the scalars through second derivatives of the one-loop potential,
\begin{equation}
    V_1''(\phi) \sim \left(\frac{\partial}{\partial \phi}m^2\right)^2 \log m^2.
\end{equation}
The Goldstones then again cause divergences, which are typically regulated away with an IR regulator inserted by hand. This IR divergence can arise because $V''$ is \emph{not} the pole mass of the particle. When calculating the pole mass, there are contributions from the momentum-dependent part of the self-energy which cancel the IR divergence in $V''$.

In my opinion, it is much simpler to use \MSbar{} and not have to deal with these issues. But there is no real inconsistency with using the mass-dependent scheme together with an IR cutoff---if the calculations are insensitive to the cutoff.
\subsection*{IR divergences}
\noindent{}Attempting to use the $\hbar$-expansion (strict loop expansion) to find the critical temperature in a theory with a radiative barrier leads to IR divergences~\cite{Laine:1994bf, Patel:2011th}. We can find the critical temperature $T_c$ as the temperature where the difference in energy $\Delta V = V(\phi_A)-V(\phi_B)$, between the two phases $\phi_A$ and $\phi_B$, vanishes. The divergence comes from expanding the critical temperature around $T_0$, the leading-order contribution,
\begin{align}
    T_c &=T_0+\hbar T_1+\mathellipsis,\\
    \left.\Delta V \right|_{T_c}&=0 \implies \left. m_3^2\right|_{T_0} =0.\label{eq:T0}
\end{align}
In the strict loop expansion $T_0$ coincides with the temperature where the classical potential of the 3D EFT has a second-order phase transition, as stated in equation~\eqref{eq:T0}. When evaluating the two-loop potential at this temperature, there are divergences $\sim \left.\log  m_3^2\right|_{T_0}$. What is going wrong?

The problem is that the wrong power counting is used. By using the modified power counting $\lambda \sim g^3$ which a radiative barrier necessitates~\cite{Arnold:1992rz}, one instead finds the critical temperature 
\begin{align}
    T_c &=T_{\mathrm{LO}}+x T_{\mathrm{NLO}}\mathellipsis,\\
    \left.\Delta V \right|_{T_c}&=0 \implies \left. \Delta V_{\mathrm{LO}}\right|_{T_{\mathrm{LO}}} =0.
\end{align}
Expanding around $T_\mathrm{LO}$ does not feature IR divergences~\cite{Ekstedt:2020abj, Ekstedt:2022zro}. Power counting together with a strict perturbative expansion is IR safe.
\subsection*{Imaginary potentials}
\noindent{}A potential with a non-zero imaginary part signals an instability: the 1-loop potential can develop an imaginary part if evaluated at a field-value where a square mass is negative. This can happen close to the broken minimum, where the Goldstone square mass changes sign. But it can also happen at the origin for temperatures below $T_0$, where the scalar mass terms are negative.

These circumstances can arise if one is mixing loop orders in $V(\phi)$---if one is not using a strict perturbative expansion. When searching field space and scanning in temperatures, this necessitates taking the real part of the potential in order to get a sensible answer. But this merely cures a symptom and does not fix the real problem: unstable modes are influencing the calculation.

But if one uses the resummed leading-order effective potential $V_{\mathrm{LO}}$ in a strict perturbative expansion, then the critical temperature and broken minimum can be found order by order. The leading-order quantities $T_\mathrm{LO}, \phi_{\mathrm{LO}}$ are found from $V_\mathrm{LO}$ and subsequent orders are evaluated there: $\left.V_\mathrm{NLO}\right|_{T_\mathrm{LO}, \phi_{\mathrm{LO}}}$. No imaginary parts develop; all modes are correctly accounted for. 
\subsection*{Mirages}
\noindent{}In the section on scale hierarchies I argued, based on power counting within the 3D EFT, that the $\mathbb{Z}_2$ symmetric pure scalar theory does not have a barrier, even though a first glance suggests it does. See~\cite[p. 7]{Arnold:1992rz} for arguments based on power counting in the original theory. It is actually known that this theory cannot have a first-order phase transition~\cite{Arnold:1992rz}: the barrier turns out to have been fictitious all along. Let us call such illusory barriers \textit{mirages}, to emphasize the danger they pose. We can contrast mirages with real radiative barriers, such as the radiative barrier of the Abelian Higgs model at high temperatures~\cite[p. 7]{Arnold:1992rz}---which arises due to a hierarchy of scales.

Generally. mirages can arise when perturbative orders are out of control. Either because the expansion parameter is too large, as above, or when orders are mixed haphazardly. So to protect ourselves against mirages we should use a consistent power counting, and perform a strict expansion in a small parameter. This also makes the $\mathbb{Z}_2$ symmetric pure scalar theory an important test case for a prospective resummation method. If a mirage can be seen in this theory, then the resummation method must be reconsidered.

\subsection*{Perturbative breakdown}
\noindent{}In the extreme case of a perturbative breakdown, mirages are indicative of a larger problem: we cannot trust the perturbative expansion.

Certain perturbative problems can be fixed by reordering perturbation theory---by resummations---while others are incurable. Famously, non-Abelian gauge theories suffer from the Linde problem: at high temperatures the gauge boson must develop a "magnetic mass" $M\sim g^2 T$ to cure IR divergences at four loops and higher~\cite{Linde:1980ts}. This results in a complete perturbative breakdown; perturbative methods cannot reach $\ordo{M^3\sim g^6 T^4}$ (four loops).

To be fair, power counting and a strict expansion cannot solve this problem. But it can dissolve it. In $\mathrm{SU}(2)$ gauge theory with a radiative barrier, the Linde problem affects the sixth order and higher in strict perturbation theory. The first five orders are calculable, and even the first three orders offer good accuracy for a wide range of expansion parameter values~\cite{Ekstedt:2022zro}. Hence, the Linde problem should not occupy too much space in our minds when studying phase transitions.
\subsection*{Resummation method dependence}
\noindent{}Comparing resummation methods and finding a difference indicates that at least one of them is wrong. But it cannot tell you which result is the correct one, or if any of them are. Instead, it is better to look at the assumptions made on which the resummations are based.

To be confident in which resummation method to use, we should instead derive it from a consistent power counting, and make sure that perturbation theory is converging. In the end, there should only be one resummation method available for a given approximation scheme: that which is implied by the hierarchy of scales.
\subsection*{Linear terms}
\noindent{}Terms linear in $\phi$ will prevent the existence of the symmetric minimum at high temperatures. Such terms contradict our usual understanding that the symmetry is restored in this limit. In the beginning of the 1990s, there were several resummation methods that produced such linear terms~\cite{Brahm:1991nh, Shaposhnikov:1991cu}, and there was some doubt whether the linear terms should exist or not.

But a convincing argument against such terms can be established from an EFT perspective, using the methods of power counting. In~\cite{Dine:1992wr}, the authors argue that the existence of an IR cutoff---the magnetic mass $\sim g^2 T$ discussed above---means that the effective potential must be analytic in $|\phi|^2$ as $\phi\rightarrow 0$. This prevents the existence of linear terms. Any consistent resummation method must respect this constraint.\footnote{Caveat: this argument only works when $\phi$ breaks a non-Abelian gauge symmetry, as only non-Abelian gauge fields have a magnetic mass. I am not aware of an argument that works for generic theories.}

In a more modern setting, it is sometimes argued that Parwani resummation can give rise to such linear terms~\cite{Croon:2020cgk}. But I think this is unfair to Parwani resummation. The linear terms arise if one uses a high-temperature approximation for the thermal counter-term, but maintains the full unexpanded integrals in the potential~\cite{Laine:2017hdk}. If the high-temperature expansion does not apply, then this is inconsistent and there will remain uncanceled terms that the counter-term insertion procedure cannot handle.

I think it is more fair to blame a faulty power counting: linear terms can arise if one is not using a consistent expansion.
\section*{Go forth and count powers}
\noindent{}I have in this paper given evidence that reshuffling the perturbative expansion is not done without risk. If one cannot establish a hierarchy of scales to motivate a resummation, then one should be wary of resumming. And if such a hierarchy of scales exists, we can then derive the corresponding resummation method.

More formally, a strict perturbative expansion derived from an EFT is an asymptotic expansion. And asymptotic expansions are \emph{unique}~\cite{bender1999advanced}: there should not be any question as to which resummation method to use once you are settled on an approximation scheme consisting of chosen degrees of freedom and a demonstrated hierarchy of scales.\footnote{The uniqueness of the coefficient also implies the perturbative expansion is gauge invariant order-by-order~\cite{Ekstedt:2022zro}.} 

In the previous sections I showcased the strength of power counting by demonstrating how it dissolves the laundry list of confusion surrounding phase transition calculations (mirages, imaginary potentials, scheme dependence, \ldots). Though I illustrated the problems using studies of phase transitions, the actual problems are quite generic, and many of them show up in other applications of QFT. 

A positive example of well-grounded resummations is the use of Soft-Collinear Effective Theory, which describes soft and collinear gluons in high-energy particle physics processes. This EFT has allowed putting direct resummation methods on a firm footing---and even enabled derivation of new methods~\cite{Becher:2014oda}. There is also more recent development in using EFTs to describe jet processes~\cite{Becher:2015hka}. 

To balance against the successes of deriving resummation methods from EFTs, we should also consider the challenges. What if there are several possible EFTs to describe the physics? Then we must indeed compare them against each other. Picking the correct EFT can be difficult, though there are a couple of approaches to making the correct selection. One can for example establish consistency of the treatment of the physical system, or one can compare with some form of data. Here nuclear physics can serve as an illustrative example. Due to the existence of many scales close to each other,  and due to the many degrees of freedom involved, finding a completely well-behaved perturbative EFT description has proven difficult~\cite{vanKolck:2020llt, Hammer:2019poc, Griesshammer:2021zzz}. There is a rich interplay of theoretical developments, feedback from experiments, and lattice calculations. 

We have a different but related situation in the study of phase transitions. It is not always clear-cut which power counting to use for a given parameter point in a particular model. This poses problems for scanning large sections of parameter space to find first-order phase transitions. Ideally, one should divide parameter space into different regions in which different perturbative expansions apply.

But how small should $\lambda$ be before we start counting it as $g^3$ instead of $g^2$? This is a question with no clear answer within perturbation theory. And unfortunately, the answer has bearing on whether a first-order phase transition occurs at all. Currently, the best we can do is to compare with results from lattice data, and to monitor the performance of the perturbative expansion. There are lattice studies available for certain simple models which capture the important dynamics of many more elaborate models, and can be used for comparison (as done in~\cite{Gould:2019qek}). Finding a possible first-order phase transition for a model that cannot be mapped to a model already studied on the lattice is a strong incentive to perform a new lattice study.

In the end, I argue that this problem is a better one to have, compared with the conceptually confusing problems on the laundry list. The question of how large different contributions are---which EFT is correct---is an honest, quantitative, and physical question. You have to be realistic about these things. 

On the other hand, much have been said about the apparent small size of uncertainty due to gauge dependence~\cite{Garny:2012cg}. For a given gauge fixing method, the results do not differ much between Landau gauge ($\xi = 0$) and reasonable values of $\xi$. It is argued that even though a gauge dependent result is uncomfortable, it is not a large quantitative issue and can be ignored~\cite{Croon:2020cgk, Schicho:2022wty}.

But I think this is a distraction from the true issue. The real problem with gauge dependence was never that one finds a span of values for different $\xi$ and have to pick one of them at the expense of accuracy. The real problem is that gauge dependence signals that something is wrong with the perturbative expansion. And when something is wrong with perturbation theory, we are at the mercy of the other problems in the laundry list.\footnote{Though to be fair, varying $\xi$ can lead not only to quantitative uncertainty, but also to qualitative differences. Sometimes barriers disappear, sometimes new minima are generated, as $\xi$ is varied~\cite{Athron:2022jyi, Ekstedt:2022zro, Zuk:2022qwx}.} Gauge dependence signals that we have lost control of perturbation theory. By not taking the signal seriously, we are bound to be confused.

The reasons above are why I encourage anyone interested in performing a perturbative study of a phase transition to always begin with the power counting, and to monitor the convergence of perturbation theory---by comparing the different orders to each other and checking that the renormalization scale dependence is under control.

Whether the high temperature expansion applies or not, you should be able to establish a hierarchy of scales and derive a consistent resummation scheme. And if no hierarchy exists, you should be wary of resumming. There are many phenomenological models in which first-order phase transitions seem possible, but the existence of which have not yet been established using power counting or nonperturbative methods.\footnote{An exercise for the reader: develop a power counting for a radiative barrier to occur in the inert doublet model.} To do so would enable us to study phase transitions in such models consistently and accurately. It would also motivate further lattice studies to better understand our perturbative expansions.

To put resummation methods on sound footing, we should be using EFTs. Go forth and count powers.

\begin{acknowledgments}
\noindent{}I thank Andreas Ekstedt and Tuomas Tenkanen for their comments on the manuscript, and Oliver Gould for enlightening discussions.
\end{acknowledgments}

%\clearpage
\appendix
\allowdisplaybreaks
\bibliographystyle{apsrev4-1}
\bibliography{references.bib}
\end{document}